\documentclass[reprint,prb,aps,superscriptaddress,showpacs]{revtex4-1}
\usepackage{epsfig,color}
\usepackage{graphicx}
\usepackage{amsmath}

\begin{document}
\title{Resonant harmonic generation and collective spin rotations\\ in electrically driven quantum dots}

\author{M. P. Nowak}
\affiliation{AGH University of Science and Technology, Faculty of Physics and Applied Computer Science,\\
al. Mickiewicza 30, 30-059 Krak\'ow, Poland}
\author{B. Szafran}
\affiliation{AGH University of Science and Technology, Faculty of Physics and Applied Computer Science,\\
al. Mickiewicza 30, 30-059 Krak\'ow, Poland}
\author{F. M. Peeters}
\affiliation{Departement Fysica, Universiteit Antwerpen, Groenenborgerlaan 171,
  B-2020 Antwerpen, Belgium}

\date{\today}

\begin{abstract}
Spin rotations induced by an ac electric field in a two-electron double quantum dot are studied
by an exact numerical solution of the time dependent Schr\"odinger equation in the context of
recent electric dipole spin resonance experiments on gated nanowires. We demonstrate that the splitting
of the main resonance line by the spin exchange coupling is accompanied by the appearance
of fractional resonances and that both these effects are triggered by interdot tunnel coupling.
We find that the ac driven system generates residual but distinct harmonics
of the driving frequency which are amplified when tuned to the main transition frequency.
The mechanism is universal for electron systems in electrically driven potentials
and works also in the absence of electron-electron interaction or spin-orbit coupling.
\end{abstract}


\maketitle
\section{Introduction}
The idea \cite{ld} of processing quantum information stored in spins of electrons confined
in quantum dots has motivated a significant theoretical and experimental effort within the last decade.
One of the necessary pre-requisites for quantum gating is coherent single spin manipulation.
Single-spin rotations can be performed using electron spin resonance  -- Rabi oscillations in external microwave radiation resonant with the Zeeman splitting of energy levels in a magnetic field ($B$). Electron spin resonance was implemented in a quantum dot \cite{esr} using an embedded on-chip microwave source.
In gated quantum dots the microwave field has been successfully replaced by ac voltages.\cite{nowack,laird,nadj-perge,brunner,schroer,nadj-perge2}
The periodic motion of the electron induced by the ac field subjects its spin to an oscillating momentum-dependent spin-orbit (SO) field,\cite{rashba, rashba2}
leading to the electric-dipole spin resonance (EDSR).\cite{edsr}
The spin rotations are detected in  two-electron double quantum dots systems \cite{esr,nowack,laird,nadj-perge,brunner,schroer,nadj-perge2} using the Pauli blockade of the current flow which occurs when the dots become occupied by electrons with parallel spins. The rotation of the spin lifts the Pauli blockade when the frequency of the ac electric field is tuned to the resonant transition.

The detailed structure of the EDSR was recently resolved \cite{nadj-perge2} in a double dot produced in a gated InSb
quantum wire with strong SO interactions. The experimental data [Fig. 2(b) of Ref. \onlinecite{nadj-perge2}] include a double line corresponding
to transitions from the spin-polarized triplet $T_+$ ground-state to a doublet formed by: 1) singlet $S$,  and 2) unpolarized triplet $T_0$, as well a single line at half the resonant frequency. Half-resonances were previously observed also in InAs quantum wire dots [Fig. 2(b) of Ref. \onlinecite{schroer}]
as well as in GaAs planar quantum dots.\cite{laird}
   Analysis of the
dipole moment induced by an ac field in the {\it singlet}-subspace of two-electron systems
was given in Ref. \onlinecite{rashbac} in terms of flopping the pseudospin mode.

In this paper we report on the solution of the time-dependent Schr\"odinger equation for the two-electron system 
in an ac field induced by gates in the presence of SO coupling.  We find spin transitions involving both Rabi oscillations for the resonant driving frequency as
well as fractional resonances which are consistent with the experimental data.\cite{nadj-perge2} We show that the mechanism responsible for the appearance
of the fractional lines is the resonant amplification of the higher harmonics residually present within the driven system.

Solution of the time-dependent Schr\"odinger equation is one of the methods \cite{awasthi,abusam,ciap} applied in theories of high harmonics generation by atoms and molecules in intense laser fields \cite{sc1,kr1,le1} in the quest for
controllable sources of ultra energetic photons. Noble gases or simple molecular systems (N$_2$, O$_2$, CO$_2$) generate {\it non-resonantly} high harmonics
of the driving laser field of intensity  $10^{11}$ W/m$^{2}$ with local field amplification by plasmonic metal nanostructures,\cite{kim}
or $10^{13}$ W/m$^{2}$ in standard conditions. The amplitude of the ac electric field applied in EDSR for quantum dots (a fraction of kV/cm) corresponds to a laser radiation of only $10^5$ W/m$^2$. Nevertheless, we find a distinct -- although residual -- appearance of a second and third harmonics of the driving
frequency $\omega_{ac}$ in the electron motion within the double dot. We demonstrate that the harmonics of the driving frequency are essentially reinforced when brought to resonance with the Rabi direct transition frequency.
We indicate that this phenomenon is quite general for ac driven electron systems confined in quantum dots, in particular that it
appears also for a single electron and in the absence of SO coupling. As a result, the confined system is driven into an excited state by  frequency $\omega$ being a fraction of the excitation energy $\Delta E$, i.e. $\hbar \omega=\Delta E/n$,
which is similar to $n$-photon optical transitions.\cite{shirley}

\begin{figure}[ht!]
\epsfxsize=70mm
               \epsfbox[57 335 580 521] {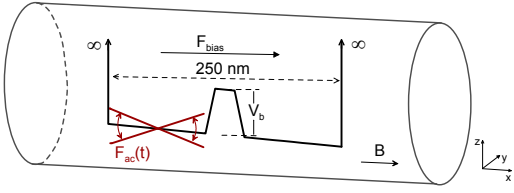}
                \caption{(color online) Schematic of the considered confinement potential of a nanowire double dot system.}
 \label{fig0}
\end{figure}
\section{Model}
The considered two-electron system is described by the  Hamiltonian
\begin{equation}
H=h_1+h_2+\frac{e^2}{4\pi\varepsilon_0\varepsilon |\textbf{r}|} ,
\end{equation}
 with the single-electron energy operator,
\begin{equation}
h_i = \frac{\hbar^2\textbf{k}^2_{i}}{2m^*} + V(\textbf{r}_i,t) + \frac{1}{2}g\mu_BB\sigma_{x_i}+ H_{SO},
\end{equation}
with magnetic field $B$ aligned along the $x$-direction. The momentum operator is $\hbar \textbf{k}_i = - i \hbar \nabla_i$ as we neglect the orbital effects of the magnetic field for low values of $B$ and in strong confinement in the plane perpendicular to the $x$-direction. The $V(\mathbf{r},t)$ stands for the confinement potential
taken in a separable form $V(\mathbf{r},t)=V_x(x,t)V_{y,z}(y,z)$.
 We include Rashba SO interaction \cite{rashba2} resulting from the electric field -- generated by the system of gates on which
 the nanowire is deposited -- which is assumed perpendicular to the wire (parallel to the $z$-direction), $H_{SO} = \alpha(\sigma_{x}k_{y}-\sigma_{y}k_{x})$.
 Figure \ref{fig0} depicts the considered confinement potential. The structure is assumed 250 nm long, with
\begin{equation}
 V_x(x,t)=V_{sx}(x)+e F_{bias} x +eF_{ac} x f(x) \cos(\omega_{ac}t) \label{kt}.
\end{equation}
  The last term in Eq. (\ref{kt}) represents the ac field, which is applied to the left dot only (see [\onlinecite{nadj-perge2}]), i.e. $f(x)=1$ in the left dot and 0 outside. $V_{sx}$ is a double quantum well potential with a $30$ nm -thick barrier of height $V_b$ in the center. A constant $F_{bias}=-0.1\;\mathrm{kV/cm}$ is taken
for $8$ mV source-drain bias voltage.\cite{nadj-perge2}
We assume a strong radial parabolic confinement in the $(y,z)$ direction which freezes the lateral wave functions of both electrons into Gaussians $\Psi=(\sqrt{\pi} l)^{-1}\exp \left[-(y^2+z^2)/2l^2 \right]$, with $l=30$ nm. Upon integration of (1) with the lateral wave functions
one arrives at an effective Hamiltonian \cite{bednarek}
\begin{eqnarray}
H&=&\sum_{i=1,2} \left[ -\frac{\hbar^2}{2m^*}\frac{\partial ^2}{\partial x^2_i}+V_x(x_i,t)-\alpha\sigma_{y_i}k_{x_i}\right.\nonumber\\
&+&\left.  \frac{1}{2}g\mu_BB\sigma_{x_i}\right] +  \frac{\sqrt{\pi/2}}{4\pi\varepsilon_0\varepsilon l}\mathrm{erfcx}\left[\frac{|x_1-x_2|}{\sqrt{2}l}\right],
\end{eqnarray}
which is used in this work.
Unless stated otherwise we apply material parameters \cite{params} for InSb: $m^*=0.014m_0$, 
 $g=-51$, $\varepsilon = 16.5$ and take the Rashba constant $\alpha=50$ meVnm.
The ac field amplitude $F_{ac} = 0.1\; \mathrm{kV/cm}$ is assumed. All the calculations are performed within the finite difference
scheme on the $(x_1,x_2)$ space with exact inclusion of the electron-electron correlation.

\section{Results}
\begin{figure}[ht!]
\epsfxsize=85mm
                \epsfbox[27 240 573 611] {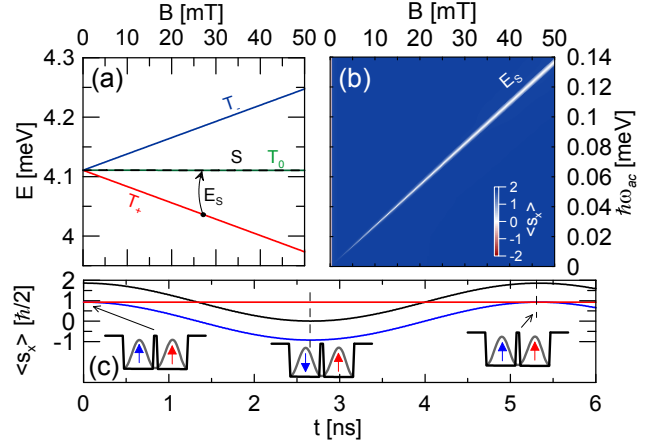}
                 \caption{(color online) Energy spectrum of coupled quantum dots with barrier height $V_b=100$ meV.
                  (b) Map of the spin transitions after $30$ ns: the minimal value of the $x$ component of the spin obtained
                  during the simulation in $\hbar/2$ units (the initial value is 2). (c)
                  Evolution of the spin $x$ component in the left (blue curve), right (red curve)
                  and both dots (black curve) at the resonance marked with the black arrow in (a).}
 \label{fig1}
\end{figure}

Figure \ref{fig1}(a) shows the energy spectrum of weakly coupled quantum dots ($V_b=100$ meV).
The four-fold degeneracy of the ground state at $B=0$ is due to weak tunnel coupling between the dots.
The degeneracy is lifted in nonzero $B$:
the triplet energy levels with spin aligned parallel and antiparallel to the $x$-direction -- $T_+$ and $T_-$ respectively \cite{note} (plotted with the red and blue curves) -- are split by the Zeeman interaction. The two other states -- $T_0$ and $S$ (plotted as the green solid and black dashed curves) -- with zero value of spin component in the $x$-direction remain degenerate (with energy separation below $0.1\;\mu$eV).

We initialize system in the $T_+$ state. For $B=20$ mT and ac frequency tuned to the energy difference between the $T_+$ and $S$ states ($\hbar \omega_{ac} = E_S \simeq 55\; \mu\mathrm{eV}$ which corresponds to the oscillation period $\tau_{ac} \simeq 75$ ps) we see
[Fig. \ref{fig1}(c)] that after about $2.7$ ns the spin of the electron in the left dot
-- wiggled by the ac field -- is inverted, while the spin in the right dot remains unaffected.

The EDSR experiments probe the spin-rotations by measuring the map of current leakage through the spin blockade as function of the driving ac frequency $\omega_{ac}$ and external magnetic field magnitude $B$.
Figure \ref{fig1}(b) shows the minimal value of the spin obtained during a time evolution
of 30 ns as function of $B$ and the driving frequency (the initial $x$ component of the spin is 2 in $\hbar/2$ units). A single line corresponding to the
$T_+\rightarrow(S,T_0)$ doublet transition is obtained.  Outside this resonant line the ac field
does not influence the spin. Note, that the
transition $T_+\rightarrow T_-$ is not observed since it requires
rotation of both spins.

\begin{figure}[ht!]
\epsfxsize=85mm
                \epsfbox[27 283 573 651] {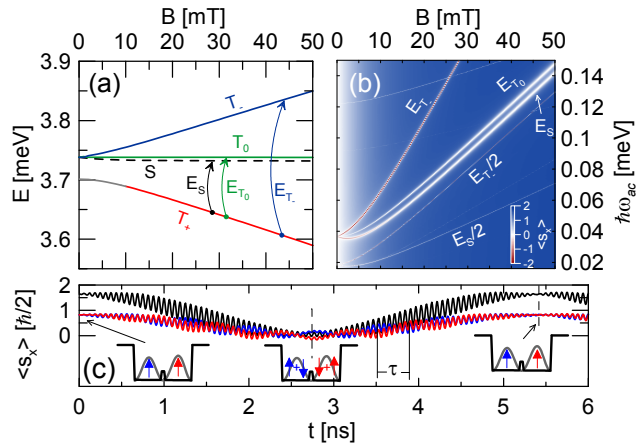}
                 \caption{(color online) Same as Fig. \ref{fig1} but for stronger interdot coupling, i.e. $V_b=17$ meV.}
 \label{fig2}
\end{figure}
For stronger interdot coupling ($V_b=17$ meV) the splitting of $T_0$ and $S$ energy levels (the exchange energy \cite{ld,sei}) becomes nonzero, $J=E_{T_0} - E_{S}\simeq5.6\;\mu$eV
[see the energy-spectrum in Fig. \ref{fig2}(a)]. For the $\omega_{ac}$ tuned to the $T_+\rightarrow S$ transition we observe
that the ac field applied to the left dot rotates the spins in {\it both} dots [Fig. \ref{fig2}(c)].
This is due to the spin-exchange interaction which is now activated by the interdot tunneling.
The characteristic spin swap time $\tau = \pi\hbar/J \simeq 370$ ps corresponds
to the intervals between the local extrema of spins observed in Fig. \ref{fig2}(c). In the plot there are also fast oscillations of the spin component visible. Their period corresponds to the period of the ac field, i.e. $\tau_{ac}\simeq63$ ps for $\hbar \omega_{ac} = 66\;\mu$eV. They are due to spin precession \cite{mpprec} induced by the spatial electron oscillation driven due to stronger interdot tunnel coupling as compared to $V_b = 100$ meV case.
Near B=0 we observe an avoided crossing of lowest-energy levels due to the SO interaction. For $B>0.1$ mT they can be identified by their spin-x component as S, $T_+$, $T_0$, and $T-$, see also the end of the Section.

The map of minimal spin states encountered during a 30 ns simulation is presented in Fig. \ref{fig2}(b).
At the diagonal of the plot two major lines emerge. They correspond to the transitions from the $T_+$ state to the $S$ (for $\hbar \omega_{ac} = E_s$) and $T_0$ (for $\hbar \omega_{ac} = E_{T_0}$) states with $0$ spin $x$ components.
 This double line was observed in the experiments \cite{schroer, nadj-perge2}
 and attributed to different $g$-factors in the dots with the assumption that a single spin responds to the ac field, and the
local differences in $g$ factor are due to variation of the confinement composition.
In fact the present simulation shows that the lines are split when $J\neq 0$, which implies the coupling between the spins in both the dots. Rotation
of the spin in the left dot to which the ac field is applied results in the spin rotation of the other electron.
 When only the spin in the left dot is inverted, the final state corresponds to a spatial ''spin density wave''
 which is not an eigenstate of the spin, but a superposition of $S$ and $T_0$ states, which can be a stationary Hamiltonian eigenstate
 only provided that $S$ and $T_0$ are degenerate as it is the case in Fig. 2.

 The transition $T_+\rightarrow S$ lifts {\it directly} the Pauli blockade, while the
 $T_+\rightarrow T_0$ transition lifts the blockade only {\it indirectly} \cite{jap} due to the mixing of $S$ and $T_0$ states -- that are close in energy -- by the hyperfine field. The red curve in Fig. \ref{fig2}(b) obtained for $\hbar \omega = E_{T_-}$ corresponds to the transition to the $T_-$ state which requires the rotation of spins in both dots, and is therefore not visible for $V_b = 100$ meV [see Fig. \ref{fig2}(b)]. In nonzero $B$ the $T_-$ energy level is too far on the energy scale to mix with the $S$ state via the nuclear spins. For that reason the transition to $T_-$ state does not unblock the current flow \cite{jap} and therefore this line is missing in the experimental data \cite{schroer,nadj-perge2} of the frequencies lifting the spin blockade of the current.

\begin{figure}[ht!]
\epsfxsize=75mm
                \epsfbox[29 245 565 600] {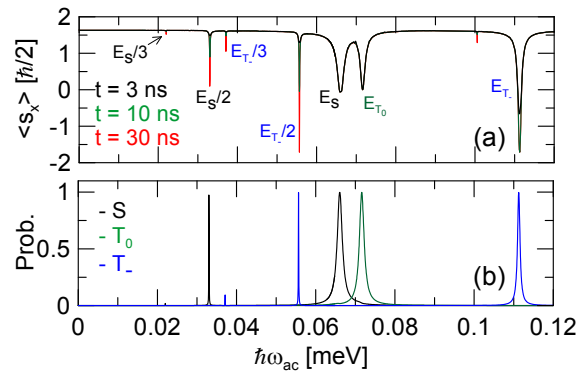}
                 \caption{(color online) (a) Spin transitions obtained for $B=20$ mT after $3$ ns (black), $10$ ns (green) and $30$ ns (red) for $V_b =17$ meV
                 (minimal $x$ component of the spin acquired by the system subjected to ac field with $T_+$ as the initial state). (b) Probability of transition to the $S$ (black curves), $T_0$ (green curves) and $T_-$ (blue curves) states after 30 ns.}
 \label{fig3}
\end{figure}

Besides the direct Rabi transitions additional ones for lower frequencies are clearly visible in Fig. \ref{fig2}(b).
 Let us focus on a cross section of the map Fig. \ref{fig2}(b) obtained for $B=20$ mT presented in Fig. \ref{fig3}(a).
  The transition probability is plotted in Fig. \ref{fig3}(b). The three broad peaks (marked with $E_S$, $E_{T_0}$ and $E_{T_-}$) correspond to direct Rabi transitions.
 The narrow resonances observed for lower $\omega_{ac}$ correspond to fractions of the frequencies of the direct transitions. The transition probabilities depicted in Fig. \ref{fig3}(b) exhibit one-half and one-third (those are not fully saturated in the plot resolution, i.e. $\hbar\Delta \omega_{ac} = 50$ neV) $T_+ \rightarrow S$ transitions (black curve) and fractional $T_+ \rightarrow T_-$ transitions (blue curve). The fractional transition to the $S$ state (for  $\hbar \omega_{ac} = E_s/2$) is of particular importance as it lifts the spin blockade in the EDSR experiments and this is the fractional resonant line that is visible in the experimental maps of Refs. \cite{laird, nadj-perge2, schroer}.
 The direct Rabi oscillations are rather slow \cite{sherman} but the fractional ones are even slower.
In Fig. \ref{fig3}(a) the transitions
 after 3 ns, 10 ns and 30 ns are plotted. For 3 ns -- the direct transitions
 are already fully resolved in contrast to the fractional ones. At the left upper corner of the map Fig. \ref{fig2}(b) one can observe additional resonance line which is a fractional resonance to the fourth excited state.

\begin{figure}[ht!]
\epsfxsize=75mm
                \epsfbox[27 238 580 614] {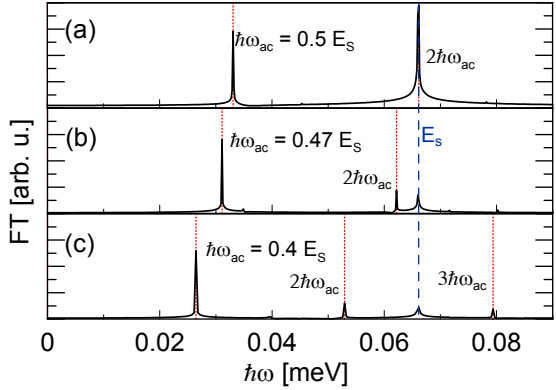}
                 \caption{(color online) Fourier transform of the total momentum calculated for three different driving ac frequencies $\hbar \omega_{ac}$.}
 \label{fig4}
\end{figure}

Let us focus on the origin of the fractional transitions. First we consider the ac frequency $\hbar \omega_{ac}$ for which no transition occurs. The Fourier transform of the total momentum is presented in Figs. \ref{fig4}(b,c).
We observe that when the electron is driven by an ac field its motion is periodic consisting of: i) the driving frequency $\hbar \omega_{ac}$, ii) its harmonics [marked with the red dashed lines in Fig. \ref{fig4}] and iii) the resonant frequency corresponding
to the direct
$T_+\rightarrow S$ transition [marked with the blue dashed line at Fig. \ref{fig4}]. When the driving frequency is such that one of its harmonics matches the resonant one [Fig. \ref{fig4}(a)] its amplitude is greatly amplified and the system exhibits a resonant transition.

We find that there is a relation between the intensity of the fractional line in the transition maps and the probability
of finding both electrons in the same dot in a given final state. In particular for $V_b=17$ meV the
probabilities for subsequent states are: $S$ -- 0.02, $T_0$ -- 0.002 and $T_{-}$ -- 0.04. Both the transitions for $\hbar \omega_{ac}=E_s/2$ and $\hbar \omega_{ac}=E_{T_{-}}/2$ occur in $t\simeq25$ ns with the half-width of the transition peak approximately $100$ neV wide (the transition for $\hbar \omega_{ac} = E_{T_{-}}/3$ is as long as $200$ ns with the half-width of the peak about $10$ neV). The fractional transitions to $T_0$ are missing in Fig. \ref{fig3} and we do not observe the generation of a residual frequency for $\hbar \omega=E_{T_0}$ in Fig. \ref{fig4}.
In fact the fractional transition for $\hbar \omega_{ac}=E_{T_0}/2$ does occur but with a line width that is narrower
than the plot resolution of $50$ neV -- the half-width of the peak is about $1$ neV with the transition time more than $t=1\;\mu$s.
For non-zero double occupancy probability the electrons are at least partially driven over
the entire double dot area by the ac field and the generation of higher harmonics becomes effective.
Consequently for $V_b=100$ meV where the probabilities are as small as $10^{-4}$ no fractional transitions are observed.

\begin{figure}[ht!]
\epsfxsize=85mm
    \epsfbox[25 305 571 547] {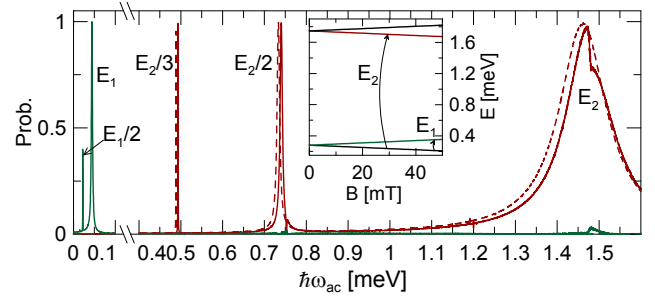}
                 \caption{(color online) Transition probability to the first-excited (green curve) and second-excited (red curve) state after $10$ ns for one-electron single dot ($V_b=0$) with (solid curves) SO coupling and without (dashed curve). Inset presents the single-electron energy spectrum with SO coupling included along with the direct Rabi transitions.}
 \label{fig5}
\end{figure}

The source of the fractional resonance observed in the experiments is the dynamics of a non-adiabatically driven electron system.
In order to demonstrate that
let us reduce the problem to a single-electron one (we also lift the interdot barrier).
 The energy spectrum for such a system is presented in the inset to Fig. \ref{fig5}. The transition between the ground state and the first-excited state is only possible through a spin rotation. The transition for $\hbar\omega_{ac} = E_2$ occurs between the states of the same spin. In Fig. \ref{fig5} we show the transition probabilities to the first- and second- excited states. We observe both the direct transitions
   and the fractional ones.
   When we switch off the SO coupling the transition to first excited state is blocked as spin becomes decoupled from the electron motion. However the transitions to the second excited state along with its fractional components is still present -- see the red dashed curve at Fig. \ref{fig5}.

\begin{figure}[ht!]
\epsfxsize=85mm
                \epsfbox[27 283 573 651] {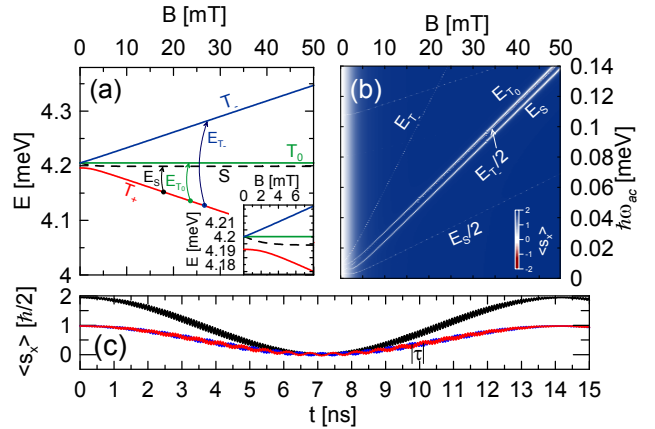}
                 \caption{(color online) Same as Fig. \ref{fig2} but for weaker SO coupling strength. The map in (b) is obtained after 60ns.}
 \label{fig6}
\end{figure}

For completeness we present the case of weaker SO coupling -- namely we apply $\alpha = 25$ meVnm.
The energy spectrum for strongly coupled dots (we chose $V_b=27$ meV to obtain similar coupling strength between $S$ and $T_0$ as previously, i.e. the exchange energy $J=6.3\;\mu$eV) is presented in Fig. \ref{fig6}(a).
For lower value of $\alpha$ one obtains a singlet and a triplet energy levels at $B=0$ and the avoided crossing between T+ and S energy level occurs at $B>0$.
Weaker SO interaction results also in a longer spin rotation time [see Fig. \ref{fig6}(c)] and now the $T_+\rightarrow S$ transition occurs in 7 ns (compared to 2.7 ns in
Fig. 3). The exchange driven small-amplitude spin oscillation with $\tau = 328$ ps are visible in Fig. 7(c) along with fast oscillation due to spin precession. In the map of spin transitions Fig. \ref{fig6}(b) both the double central line and the half-frequency transitions to the S and $T_-$ states appear. Only now the lines are narrower as compared to stronger SO interaction case (as the amplitude of the SO effective magnetic field that drives the spin transition is decreased). The half-frequency transitions occurs in 54 ns (compared to 25 ns for $\alpha = 50$ meV nm). Otherwise the EDSR transition map remains qualitatively unchanged.

\section{Conclusions}
In conclusion we studied the electrically induced transitions between the electron states in quantum dots.
We presented that the electron oscillations induced by an ac field is accompanied by residual harmonics of
the driving frequency. We demonstrated that the fractional transitions observed in EDSR experiments
involve resonant amplification of the harmonics in the electron dynamics when they match the Rabi transition frequency.
Moreover we indicated that the resonant amplification of higher harmonics is an intrinsic phenomenon of a driven electron system which occurs
also for a single charge and without SO coupling.

In the two-electron system of the double dot when the ac field is only applied to one of the dots
a non-zero interdot tunnel coupling is necessary for the fractional transitions to appear as it triggers the motion of both electrons.
A consequence of the non-zero exchange energy is the splitting of the main resonance line to $T_0$ and $S$ final states.
Thus the appearance of the double resonant line
and the fractional resonance have a common origin.

\section*{Acknowledgements}
This work was supported by the funds of Ministry of Science and Higher Education (MNiSW) for 2012 -- 2013, and by PL-Grid Infrastructure. M.P.N. gratefully acknowledges the support from the Foundation for Polish Science (FNP) under START and MPD programme co-financed by the EU European Regional Development Fund.


\begin{thebibliography}{00}
\bibitem{note} Note that in the presence of SO interaction the states do not have well defined spins -- we name them  ($S$, $T_+$, $T_0$, $T_-$)  after their character without SO interaction.
\end{thebibliography}
\end{document}